\documentstyle{article}
\newcommand{\eqnsection}{
\renewcommand{\theequation}{\thesection.\arabic{equation}}
\makeatletter
\csname $addtoreset\endcsname
\makeatother}
\begin{document}
\eqnsection
\title{ Classical  integrable lattice models
 through  quantum group related formalism
}
\author{
Anjan Kundu  \\
Fachbereich 17--Mathematik/ Informatik, \
GH--Universit\"at Kassel \\
Holl\"andische Str. 36 \, 34109 Kassel, Germany
\\{\it Permanent address}:\\
 Saha Institute of Nuclear Physics, AF/1 Bidhan Nagar,
 \\ Calcutta 700 064,India.
}
\date{}
\maketitle
\begin{abstract}
%------------------------------------------------------------
We translate  effectively our earlier quantum constructions to the
classical language and using
 Yang-Baxterisation of the Faddeev-Reshetikhin-Takhtajan
 algebra are able to
 construct Lax operators and associated $r$-matrices
of classical integrable  models.
Thus new as well as known lattice systems of different classes are generated
including   new types of  collective integrable models  and canonical models
with nonstandard $r$ matrices.
\end{abstract}
%%%%%%%%%%%%%%%%%%%%%%%%%%%%%%%%%%%%%%%%%%%%%%%%%%%%%%%%%%%%%%%%%%
%\setcounter{equation}{0}
\section  { Introduction}

   The basic aim of present investigation is to
show that some effective formalism
 developed around quantum algebra and quantum
integrable systems
 can also be fruitfully applied to the domain of classical
   models. In particular we are able to translate most of our
  results related to the generation of quantum integrable models [1]
 to the {\it classical} language and  present a systematic construction
  of the Lax  operators $L(\lambda)$ and related classical $r(\lambda,\mu)
 $-matrices
 for new as well as existing  lattice models with canonical Poisson-bracket
 structures. It demonstrates that
   there exist some fundamental {\it building blocks} for both Lax operators
   and  $r$-matrices, out of which these objects can be easily built up
   following a classical analog of the Yang-Baxterisation of the
    Faddeev-Reshetikhin-Takhtajan  algebra [2].

We also show that applying analogous {\it quantum} constructions
 one can  recover cheaply though methodically many classical results,
 which were possibly discovered originally using deep intuition.
Such examples are the famous Ablowitz-Ladik  [AL] model [3],discrete time
  Toda
chain  [DTTC] [4], asymmetric  lattice NLS model [5] etc.. Moreover one can
 construct now exactly integrable
 descrete version of derivative NLS and
massive Thirrring models, where earlier attempts  failed.

Moreover the construction indicates the existence of
local as well as global {\it collective} integrable models formed by
 collecting  a number of constituent models belonging
 to the same {\it descendant} class, a possibility which
 has  been ignored mostly
in lattice model construction. Another new aspect
is the formulation of a special class of integrable systems having
nonstandared  $r$-matrices with
 both additive and difference dependence on spectral parameters.

Hopefully the results presented here
 would draw attention of {\it classical} mathematical
physicists and thus serve as a bridge narrowing the existing gap between
the two camps: classical and quantum.

\section { Building blocks and basic PB relations}

Let us consider first  simple constant solutions of the classical
Yang-Baxter equation:

\begin{equation}
  [r_{ 12},~r_{ 13}]~+ [r_{ 12},~r_{ 23}]~+ [r_{ 13},~r_{ 23}]~= 0
\end{equation}
%(2.1)
given as
$$
r^{+} = \alpha \left( \begin{array}{c}
  {1} \quad   \quad  \quad   \\
    \quad \quad   {2}  \quad   \\
      \quad \quad   \quad  \quad   \\
 \quad   \quad \quad   {1}
          \end{array}   \right), \quad
 \mbox{ and}
 \quad
r^{-} = \alpha\left( \begin{array}{c}
  {1} \quad   \quad  \quad   \\
    \quad \quad   \quad  \quad   \\
      \quad {2}  \quad  \quad   \\
 \quad   \quad \quad   {1}
          \end{array}   \right),  $$
   where parameter $\alpha$ may be taken as the {\it deformation}
   parameter    in the classical case.
    Similarly consider $L^{\pm}$ matrices also given in the
 upper/lower triangular form as
 $$
L^{+} = \left( \begin{array}{c}
  {\tau_1^+} \quad    {\tau_{21}} \\
    \quad \quad   {\tau_2^+}
          \end{array}   \right), \quad
\quad L^{-} = \left( \begin{array}{c}
 {\tau_1^-} \quad  \quad   \\
  {\tau_{12}} \quad {\tau_2^-}
          \end{array}   \right) \quad,
 $$
 with elements $\{ \tau \}$ being as yet
 undefined  dependent  variables.
As we  see below the matrices   $r^{\pm}$
 will serve as  our building blocks for the construction of
 spectral parameter dependent classical
 $r(\xi,\eta)$-matrix, while    $L^{\pm}$  will do the same
  for the   related Lax operators
    $L(\xi)$ of the integrable lattice
  models.
  To specify now the nature of  $\{\tau\}$  variables
these two sets of matrices are linked through the
   Poisson bracket (PB) relations
\begin{eqnarray}
 \{~L^{\pm} \otimes,L^{\pm} \} &=& [ r^{\pm},L^{\pm} \otimes L^{\pm}]
 ,  \\
  \{~L^{+} \otimes,L^{-} \} &=& [ r^{+},L^{+} \otimes L^{-}]
,\end{eqnarray}
%2.2-3
which are nothing but the classical analogs of the
   Faddeev-Reshetikhin-Takhtajan  algebra, a well known relation
   in the subject of quantum group.
The  relations (2.2-3)  in the elementwise form yields the defining
PB relations among the ${\bf \tau}$ variables as
\begin {eqnarray}
\{ \tau_{12}, \tau_{21} \} ~=~
2\alpha \left (  \tau_1^- \tau_2^+  -  \tau_1^+ \tau_2^- \right )
 ~, \quad \    \{\tau_i^{\pm },\tau_{j}^{\pm}\}=0, \\
\{\tau_i^{\pm },\tau_{12}\} ~=~ \pm \epsilon_i
\alpha \tau_{12} \tau_i^{\pm }~,~\quad \
\{\tau_i^{\pm },\tau_{21}\} ~=~\mp \epsilon_i
\alpha \tau_{21} \tau_i^{\pm }~,~
\end {eqnarray}
%(2.4-5)
for $i,j=(1,2)$ with $\epsilon_1=1,\ \epsilon_2=-1$.
 Interestingly, when
the {\it deforming} parameter $\alpha \rightarrow 0$, a consistent
  limit of (2.4-5) exists with
  \begin{equation}
    (\tau_i^{+ } + \tau_i^{-})
  \rightarrow K^0_i, \quad  (\tau_i^{+ } - \tau_i^{-})  \rightarrow
  \frac {1}{\alpha} K^1_i , \quad \tau_{ij}  \rightarrow K_{ij}
  \end{equation}
  % 2.6
yielding   the PB relations
\begin {eqnarray}
\{ K_{12}, K_{21} \} ~=~
   K_1^0 K_2^1  -  K_1^1 K_2^0
 ~, \quad \    \{K_1^{0 }, K_{2}^{0}\}=0, \\
\{K_i^{0 },K_{12}\} ~=~ \epsilon_i
 K_{12} K_i^{1 }~,~\quad \
\{K_i^{0 },K_{21}\} ~=~ - \epsilon_i
 K_{21} K_i^{1 }~,
\end {eqnarray}
%(2.7-8 )
where $K_i^{1}, \ i=1,2$ becomes  central elements with trivial PB
with all others.
The above Poisson algebras (2.4-5) and (2.7-8), as we  show below,
 will  play a decisive role
in generating two different  large classes of integrable models.

\section {Lax operators and $r$-matrices through  Yang -  Baxterisation}
\setcounter {equation}{0}

Equipt  with all the {\it building materials}: $L^{\pm},r^{\pm}$ matrices
and the  relations (2.4-5) or (2.7-8) we can  start  constructing
the Lax operator $L(\xi)$ and the $r(\xi,\eta)$-matrix by {\it stiching}
together
the  upper and lower triangular
 matrices through spectral parameters $\xi, \eta$ as
   \begin{equation}
   r (\xi,\eta ) ~=~ f(\xi_{12})[ \xi_{12}^{-1} r^+ +
      \xi_{12} r^-  ]~,\ \ \ \
 L (\xi)  ~=~ { \xi^{-1}}  L^{+} + \xi L^{-}
,   \end{equation}
   %3.1
 where $\xi_{12}=  \frac{ \xi}{\eta} $ and  $ f(\xi_{12})=\frac {\alpha}
 {2(\xi_{12}^{-1}-\xi_{12})}$ is a function with a pole at $\xi_{12}=1. $
To  convince ourselves that
 the $r$ and $L$ matrices thus formed can be considered
as the  representatives of  lattice integrable  models,
we should check that they satisfy the classical Yang-Baxter equation (CYBE)
\begin{equation}
 \{L_n (\xi) \otimes,L_m (\eta) \} = \delta_{mn}
 [ r(\xi,\eta),L_n(\xi) \otimes L_n(\eta)]
.\end{equation}
%3.2
This checking however becomes easy due to the  algebra (2.2-3)
along with the assumption of vanishing PB at
 different lattice points (ultralocality condition)
and the obvious relation        $r^{+}+r^{-}= 2\alpha {\bf P}  \ $
           through the permutation operator    ${\bf P}$ ( the
            classical analog
           of the Hecke condition ).
  Therefore   one may get from (3.1)
   (after an irrelevant gauge transformation) a genuine
Lax operator and the associated $r$ matrix in  explicit
form as
\begin {equation}
L(\xi) = \left( \begin{array}{c}
\xi {\tau_1^-} + \frac {1}{\xi}{\tau_1^+} \qquad    {\tau_{21}} \\
   {\tau_{12}} \qquad \xi {\tau_2^-} +\frac {1}{\xi} {\tau_2^+}
          \end{array}   \right),
\end {equation}
%(3.3)
and
\begin {equation}
r(\xi,\eta) = \left( \begin{array}{c}
a(\xi,\eta) \quad \ \quad \ \ \ \quad \\
    \quad \ \quad \ \ \  b(\xi,\eta)  \ \quad  \\
     \quad \           \
     b(\xi,\eta) \ \  \quad \ \quad \\
        \quad \ \quad \ \ \ \quad  a(\xi,\eta)
          \end{array}   \right),
\end {equation}
%3.4
where $a(\xi,\eta)  =\frac{\eta^2+\xi^2} {2(\eta^2-\xi^2)}$ and $
                    b(\xi,\eta)= \frac{\eta \xi} {\eta^2-\xi^2} $.
Note that expressing  spectral parameters in the form
$\xi=e^{-\alpha \lambda}, \  \eta=e^{-\alpha \mu}$ the dependence of the
above $r$-matrix (3.4) on trigonometric functions and moreover
 only on the defference of   parameters
as $r(\lambda-\mu)$ becomes obvious.
Remarkably at   {\it deformation } parameter $\alpha \rightarrow 0 $
 due to the existing limit (2.6) of the
  dependent variables and the expansion
 of spectral parameter  $\xi\approx 1-\alpha \lambda$
 the $L(\xi)$ operator   (3.3) reduces consistently to
       \begin {equation}
L(\lambda) = \left( \begin{array}{c}
 {K_1^0} + \lambda {K_1^1} \qquad    {K_{21}} \\
   {K_{12}} \qquad  {K_2^0} +\lambda {K_2^1}
          \end{array}   \right).
\end {equation}
%(3.5)
At the same time due to $f(\xi_{12}) \rightarrow
\frac {\alpha}{2(\lambda-\mu)}$, through  the use of the  Hecke condition
 trigonometric $r$-matrix (3.4)
reduces to  its rational limit
\begin {equation}
r(\lambda-\mu)=\frac{{\bf P}}{\lambda-\mu}
.\end {equation}
%(3.6)

We show in the next section that the Lax operator (3.3) along with
the PB relations (2.4-5) between its elements serves
 as an
excellent {\it ancestor } lattice model
 generating a large class of {\it descendants},
all sharing the same trigonometric $r(\xi,\eta)$-matrix (3.4). Similarly
(3.5) with (2.7-8) is responsible for another {\it descendant} class
having the rational form (3.6) for the associated $r$-matrix.

\section { Classical integrable lattice systems as descendant models  }
\setcounter {equation}{0}
         The idea is to insert the Lax  operators on a lattice
with $n=[1,N]$ sites
  and find consistent reductions of the general $L$ operator
         (3.3) with PB relations (2.4-5) or of (3.5) with (2.7-8). For such
         reductions proper  change of dependent variables from
         $\tau_n$ or $ K_n$ to canonical ones :
         $$\{u_n,p_m\}=\delta_{nm} \ \ \
 \mbox {or} \ \  \ \{\psi_n,\psi_n^\dagger\}
=i\delta_{nm} $$
or to some other physically interesting variables ( like in the AL model),
would result different classical integrable  lattice systems, since
by construction these descendant models would be associated with
$r$-matrix (3.4) or (3.6) and satisfy CYBE (3.2).

It is not difficult to show that symmetric reduction of the form
\begin {equation}
\tau^+_1 =(\tau^+_2)^{-1}=-\tau^-_2=-(\tau^-_1)^{-1}, \qquad \tau_{12}=
\tau_{21}^\dagger
.\end {equation}
%(4.1)
     reduces PB relations (2.4-5) to the classical analog of the $q$-deformed
     algebra $U_q(su(2))$ [7], while similar symmetric reduction
     of (2.7-8) as
\begin {equation}
K^1_1 = K^1_2=1,  \ \ K^0_1=-K^0_2, \qquad K_{12}=
K_{21}^\dagger
\end {equation}
%(4.2)
recovers  the $su(2)$ algebra.
The reduction (4.1) in turn, expressed through canonical variables $(u, p)$ ,
yields the lattice sine-Gordon model [8], while  reduction (4.2)
realised in $(\psi, \psi^\dagger)$ gives
           the standard lattice NLS model [9]. It should be remarked that
though
           such symmetric reductions are important due to their direct
           relation with  popular algebraic structures, the
significant feature of our $L$ operators (3.3) and (3.5) is
 indeed in their explicit
asymmetric form. And this in particular  allows to yield (without going
through any limiting proceedure)
the Lax operators of models  exhibiting lesser symmetry.

An important example of  such asymmetric models
 related to  (3.3) is the Lattice Liouville model with
$$\tau^-_1=\tau^+_2=0, \ \ \tau^+_1=\tau^-_2   $$
realised through $u, p$ [1]. On the other hand asymmetric reduction
 of (3.5) given by
\[ K_1^1=i, \ \ K^0_2=K_2^1=0
\]
with the expression for other functions as
       \begin {equation}
K^0_1 =p,\ \  K_{12}= (K_{12})^{-1}=e^u ,
\end {equation}
%(4.3)
yields directly the Toda lattice model.

Remarkably, some recently proposed discrete integrable systems [5,10]
can also be derived consistently as other possible asymmetric
reductions of (3.5). For example,

\begin {equation}
K^1_2=0,             \ K^0_2=K^1_1=1,     \ K^0_1=\phi_n \psi_n, \
K_{12}=\psi_n, \ K_{21}= \phi_n
\end {equation}
%(4.4)
with $n= [1,N]$ generates [11] the simple lattice model of [5]
, while
\begin {equation}
K^1_2=0,             \ K^0_2= \gamma, \ K^1_1=1,     \ K^0_1=\psi_i^* \psi_i
+ \omega_i, \
K_{12}=\psi_i, \ K_{21}= \gamma \psi^*_i
,\end {equation}
%(4.5)
with $i=1,2$ constructs the Toda-like lattice system considered in [10].
Evidently there exist  diffrent other reductions of (3.3) and (3.5)
capable of generating other integrable systems. We also get as a bonus
simultaniously the Lax operators and the $r$ matrices of the constructed
models ensuring their integrability.

It is interesting to note that such asymmetries can be made more explicit
by considering models with  new $r$-matrix solution
\begin {equation}
\tilde r= r_0+2f
,\end {equation}
%(4.6)
where $r_0$ is the original solution (3.4) or (3.6) and
\begin {equation}
f= diag(\eta'-\xi', \eta'+\xi', -(\eta'+\xi'), -(\eta'-\xi')
)\end {equation}
%(4.7)
with {\it colour} parameters $\eta', \xi'$. The expression (4.6-7)
as a new solution of (2.1) is a classical statement of the twisting
transformation  of [12], which can also be checked otherwise by direct
insertion.
We consider the particular case
\begin {equation}
\eta'=c\eta+\alpha,  \ \ \xi'=c\xi+\alpha
\end {equation}
%(4.8)
with $c$ being a constant parameter, which through
(4.6-8)  yields a new type of $r$-matrix with sum as well as difference
dependence on spectral parameters. Observe that at $c=0$, when $f= diag
(0, \alpha,-\alpha,0)$,  \  with (3.4) as $r_0$ one recovers from (4.6)
exactly  the
$r$-matrix associated with the AL  [3] as well as the  DTTC [4] models.
 A natural
expectation is therefore that these models somehow must be hidden
in our construction. To show that this is indeed the case,
we note that for this transformed $r$-matrix and with the $L$ operator
taken in the form (3.3) the PB relations are changed from (2.4-5)
consistently to
\begin {eqnarray}
\{ \tau_{12}, \tau_{21} \} ~=~
2\alpha \left (  (\tau_1^- \tau_2^+  -  \tau_1^+ \tau_2^-) + \tau_{12}\tau
_{21} \right )\\
\{\tau_i^{+},\tau_{12}\} ~=~2 \epsilon_i
\alpha \tau_{12} \tau_i^{+ }~,~\quad \
\{\tau_i^{+ },\tau_{21}\} ~=~ -2 \epsilon_i
\alpha \tau_{21} \tau_i^{+ }~,~
\end {eqnarray}
%(4.9-10)
with functions $\tau^-_i$ having now trivial PB with all other
elements. The explicit asymmetry of (4.9-10) is obvious,
which for
\begin {equation}
 \tau_1^- = \tau_2^+ =0, \   \tau_1^+ =\tau_2^-=1 , \
\tau_{12} ~=~ b_n, \
\tau_{12} = b_n^*
\end {equation}
with PB $\ \
\{b_{m },b^*_{n}\} ~=~-2
\alpha (1+ b_nb^*_n)\delta_{mn} , \ \
$yields the same AL model. On the other hand
 reduction of (4.9-10) as
\begin {equation}
 \tau_1^- =-1, \ \tau_2^{\pm} =0, \   \tau_1^+ = e^{\alpha p_n}
 , \
\tau_{21} ~=~ \alpha e^{q_n}, \ \
\tau_{12} =-\alpha e^{-q_n+\alpha p_n}
\end {equation}
gives the DTTC,  meeting our expectation.
\section {Collective models and models with nonadditive $r$-matrices}
\setcounter {equation}{0}

An interesting possibility of constructing {\it collective} models
by joining the constituent integrable elements can be effectively
exploited thanks to a symmetry of the CYBE (3.2). In particular, it is
easily checked that if $L(\xi,\tau)$ and $\tilde L(\xi,\tilde \tau)$
are Lax operators corresponding to two independent descendant models
sharing the same $r$-matrix and with $\{L,\tilde L\}=0$, then the
{\it collective} model $L(\xi,\tau,\tilde \tau)=L(\xi,\tau)\tilde L(\xi
,\tilde
\tau)$  will also be integrable with the same $r$-matrix.
This symmetry allows us to construct integrable models by collecting
similar $L$ operators at each lattice point. A recent construction [10] of
Toda-like models by joining several bosonic systems is an example of such
collective models. A more exciting example is possibly the construction
of lattice massive Thirring model by joining integrable discrete derivative
NLS models [13]. The discrete derivative NLS in turn may be given by the $L$
operator with  reduction of (3.4) as
\begin{eqnarray}
\tau^+_1&=&(\tau^+_2)^{-1}=q^{-N_n}, \ \tau^-_1=\frac{a }{4i}\ q^{N_n+1}, \
\tau^-_2=-\frac{a}{4i} \ q^{-(N_n+1)},\\
\tau_{12} &=&  (\frac {a}{2})^{\frac{1}{2}} A_n=\tau_{21}^*
, \ \ q=e^\alpha
\end{eqnarray}
where $a$  is the lattice constant, $N_n= a \psi^*_n\psi_n, $
and $A_n$ represents a $q$-boson [14] expressed as $A_n= \psi
_ng(N_n), \ \ g^2= \frac {[2N_n]_q}{N_n}$.
Using the invariance of (2.4-5) under the exchange of variables
$$ \tau^+_1   \rightleftharpoons \tau^-_2, \ \tau^+_2
 \rightleftharpoons \tau^-_1 $$
a similar operator $L^{(2)}$
with another independent component of $\psi$ variable is obtained,  yielding
 the collective model $L_n=L_n^{(1)} L_n^{(2)}
,\ $  which is the discrete version of the massive Thirring model.

Extending this idea to more global level we may even insert  completely
different descendant models at different lattice nodes, preserving only the
periodic structure. It seems that such rich possibilities for constructing
new classical integrable lattice systems with canonical
structures remain mostly unexplored.

Finally we come to another novel construction of a class of lattice models
with nonstandard $r$ matrices.
 For this we have to
choose nontrivial $c$
in (4.8), while as $r_0$ in (4.6) one may take either the trigonometric  (3.4)
or the rational form (3.6). The function $f$ naturally brings in
new ( sum as well as difference)
 dependence on spectral parameters. The associated $L$ operator may also
be
changed accordingly keeping  the PB relations (2.4-5) or (2.7-8)
unchanged. Or alternatively , the $L$ operator may again be chosen as
(3.3) or (3.5) and get the deformed PB relations from
the CYBE.
Following the first choice  we get
[1] \begin {equation}
 L(\xi) = (\tau^-_1 \tau^+_2)^{c\lambda+\alpha}\ L^0(\xi),\end {equation}
 $L^0$ being the
original $L$ operator (3.3)
with the trigonometric
 $r$ matrix. As has been shown above for $c=0$ this
system would
contain the AL and the DTTC models. Therefore in the general case it
should yield new generalised integrable models with $r$ matrix having
an interesting nonadditive  dependence on spectral parameters
and at the same time it would naturally recover all the other classes of
integrable models discussed here at different particular cases.
Similar construction exist equally  for models with rational $r$ matrix.

\section {Conclusion}
Application of quantum group related constructions to classical integrable
discrete systems yields intriguing results in model construction. One gets
the Lax operators and associated $r$ matrices in a systematic way, for
which to our knowledge no other  simple prescription exists. This also
gives the  construction of some collective integrable models as well as
models with nonstandard $r$-matrices. Such possibilities are worth
exploring for generating new type of integrable lattice models.
\\ \\ \\
\noindent {\bf Acknowledgement}\\ \\
The author extends his sincere thanks to the organisers of this conference
and the Alexander von Humboldt Foundation for support.\\
\\ \\ \noindent {\bf
References}:\\
               \\
\ [1] \ A. Kundu  and B. Basumallick \   \
 1992
 {\it Mod. Phys. Lett.  }
 {\bf A 7 }  61 ;

\ \ \ \ A. Kundu and B. Basumallick
, {\it Coloured FRT algebra and its Yang-Baxterisation
leading to integrable models with nonadditive $R$-matrix}
 Saha Inst. preprint , April 1993
\\
\ [2] \ N.Yu.
Reshitikhin , L.A. Takhtajan  and L.D. Faddeev  1989 {\it Algebra
and Analysis } \ {\bf 1} 17 \\
\ [3] \
\ M.J. Ablowitz and J.F. Ladik  1976 {\it Stud.Appl.Math.} {\bf 55}
 213 \\
\ [4] \
 Yu.B.  Suris   1990 {\it Phys.Lett.} {\bf A 145} 113\\
\ [5] \
I. Mirola, O. Ragnisco and T.G. Zhang {\it A novel hierarchy of
integrable lattices}, Rome Univ. preprint  n. 976, 1993
\\
\ [6] \ V.F.R. Jones 1990 1993 {\it Int. J. Mod. Phys.}  {\bf B4} 701
\\
 \ [7] \ M. Jimbo 1985 {\it Lett. Math. Phys.}  {\bf 20} 331;

 \ \ \ \  V.G. Drinfeld 1986 in {\it Proc. ICM (Berkeley)}  798 \\
\ [8] \ A.G. Izergin and V.E. Korepin 1982 {\it Nucl. Phys.} {\bf B 205
 [FS 5]} 401 \\
\ [9]  A.G. Izergin and V.E.Korepin 1981
{\it Sov.Phys.Dokl.} {\bf 259 } 76 ;

\ \
\ \ V.O. Tarasov, L.A. Takhtajan and L.D. Faddeev 1983
{\it Teor. Mat. Fiz.} {\bf 57 } 163
\ [10]\ P.L. Christiansen, M.F. Jorgensen and V.B. Kuznetsov,
   1993 {\it Lett. Math. Phys.}  {\bf 29} 165\\
\ [11] \  O. Ragnisco and A. Kundu
 {\it A simple lattice variant of NLS equation and its
deformation with exact quantum solution.}, Saha Inst. preprint SINP/TNP/92-15
, 1992\\
\ [12] \ N. Reshetikhin 1990 {\it Lett. Math. Phys.}  {\bf 20} 331\\
\  [13]\ A. Kundu and B. Basumallick
  1993 {\it J. Math. Phys.}  {\bf 34} 1052 ;
\\  \
 \ \ \  B. Basumallick and A. Kundu 1992 {\it  Phys.Lett.  }  {\bf B 287 } 149
\\   \ [14] \ A.J. MacFarlane  1989 {\it J. Phys.} {\bf A22} 458l

\end{document}